\documentclass[letterpaper,11pt]{article}
\usepackage{jheppub}
\usepackage{amsmath}
\usepackage{bm}
\usepackage{amssymb}
\usepackage{listings}
\usepackage{multirow}
\usepackage{makecell}
\usepackage{array}
\usepackage{subfigure}
\usepackage{siunitx}
\usepackage{booktabs}
\usepackage{amssymb}
\usepackage{pifont}
\usepackage{listings}
\usepackage{xcolor}
\usepackage{float}
\usepackage{slashed}

\usepackage{tikz}
\usepackage{pgffor}							

\newcommand{\be}{\begin{equation}}
\newcommand{\ee}{\end{equation}}
\newcommand{\beq}{\begin{equation}}
\newcommand{\eeq}{\end{equation}}
\newcommand{\bea}{\begin{eqnarray}}
\newcommand{\eea}{\end{eqnarray}}
\newcommand{\besp}{\begin{equation}\begin{split}}
\newcommand{\eesp}{\end{split}\end{equation}}

\newcommand{\Dfbd}{\mathord{\buildrel{\lower3pt\hbox{$\scriptscriptstyle\leftrightarrow$}}\over {D}_{\mu}}}

\def\0{\textbf{0}}
\def\1{\textbf{1}}
\def\2{\textbf{2}}
\def\3{\textbf{3}}
\def\4{\textbf{4}}
\def\5{\textbf{5}}
\def\6{\textbf{6}}
\def\7{\textbf{7}}
\def\8{\textbf{8}}
\def\9{\textbf{9}}

\begin{document}

\title{Size Dependence of the Sommerfeld Enhancement for Puffy Dark Matter}

\author[a]{Wu-Long Xu}
\author[b,c]{,~Jin Min Yang}
\author[d,1]{,~Wen-Na Yang,\note{Corresponding author.}}

\affiliation[a]{School of Integrated Circuits, Chengdu Technological University, Chengdu 611730, P. R. China}
\affiliation[b]{Center for Theoretical Physics, Henan Normal University, Xinxiang 453007, P. R. China}
\affiliation[c]{Institute of Theoretical Physics, Chinese Academy of Sciences, Beijing 100190, P. R. China}
\affiliation[d]{Department of Physics and Institute of Theoretical Physics,
	Nanjing Normal University, Nanjing 210023,  P. R.  China}

\emailAdd{396440567@qq.com}
\emailAdd{jmyang@itp.ac.cn}
\emailAdd{wennayang@njnu.edu.cn}

\abstract{
We examine the size effects in the Sommerfeld enhancement factor for puffy dark matter annihilation.
First, we use the partial-wave method to study the case of puffy dark matter for which only a charge density distribution is given without specifying its internal structure. We find that by using two dimensionless parameters, we can provide a characterization of the resonance structure of the Sommerfeld enhancement. Using this approach, we demonstrate that 
the finite size of dark matter particle is another fundamental factor, in addition to low velocity, that affects the Sommerfeld enhancement. Then, as an example of puffy dark matter with nontrivial internal structures, we perform the analysis for the nugget-type dark matter, whose Sommerfeld enhancement factor is found to exhibit a resonant behavior similar to that of point-like particles.}

\maketitle


\section{Introduction}\label{sec1}
Dark matter constitutes more than 80\% of the matter content of the Universe, and the evidence for its existence comes primarily from astronomical and astrophysical observations~
~\cite{Clowe:2006eq,Randall:2008ppe, Planck:2018vyg}. From the perspective of particle physics, however, no experimental signal of the most widely studied dark matter candidate, the Weakly Interacting Massive Particle (WIMP), has yet been observed in laboratory experiments
~\cite{Goodman:1984dc,Jungman:1995df,Bertone:2018krk}. Conventional direct-detection techniques are typically sensitive only to nuclear recoil energies at the $\rm keV$ scale~\cite{Lewin:1995rx,Schumann:2019eaa}. Consequently, probing lighter dark matter particles requires alternative approaches. One possibility is to increase the target mass, although this strategy may eventually encounter the limitations imposed by the neutrino fog or the neutrino floor~\cite{Akerib:2022ort,GlobalArgonDarkMatter:2024wtv,Billard:2013qya,OHare:2016pjy}. Another approach is to increase the exposure time, as in the case of paleo-detectors~\cite{Acevedo:2021tbl,Chu:2026skp,Wang:2026you}. A third possibility is to exploit mechanisms that accelerate dark matter particles, such as interactions involving cosmic rays (CRs) or the Sun, thereby endowing them with larger kinetic energies and enhancing their detectability~\cite{Bringmann:2018cvk,Su:2020zny,Ge:2024cto,Xia:2024ryt,Yang:2025ejt, Xia:2021vbz,Wang:2023wrx,PandaX:2023tfq}.

Given our limited knowledge about the fundamental properties of the dark matter particle, such as its mass, size, spin, and internal structure, it is worthwhile to explore the possibility that it may possess a finite spatial extent. If it is not point-like but instead has a nonzero size, an important question is to what extent finite-size effects can alter the phenomenology of conventional point-like dark matter. For instance, in direct-detection experiments, finite-size effects may influence the scattering of dark matter off target nuclei~\cite{Xu:2025xaq}. Quantifying these effects is essential for understanding how the predictions of finite-size dark matter models differ from those of standard point-particle scenarios~\cite{Wang:2023xgm,Wang:2021tjf}. Moreover, it is natural to ask whether the existing constraints from direct-detection experiments, indirect-detection observations, and cosmological measurements can be used to exclude some finite-size dark matter models, or at least constrain some specific candidates with distinct internal structures, such as dark matter nuggets, dark atoms, and dark glueballs~\cite{Qi:2026gan,Kim:2025uzz,Cline:2021itd,Beneke:2024iev,Li:2026nse,Carenza:2024avj}. These questions constitute one of the primary motivations for the present work.

The most widely studied dark matter productions in the early Universe are freeze-out and freeze-in mechanisms ~\cite{Baer:2014eja,Hall:2009bx}. In the standard freeze-out scenario, the relic abundance is controlled mainly by the annihilation rate of dark matter particles, whereas the freeze-in production is governed by feeble interactions that populate the dark sector from the thermal bath. In the standard WIMP paradigm, for example, the annihilation cross section of a pair of dark matter particles cannot always be adequately described by the Born approximation. At low velocities, repeated exchange of a light mediator can distort the incoming two-body wave function and significantly enhance the annihilation rate. This phenomenon is known as the Sommerfeld enhancement~\cite{Sommerfeld:1931qaf,Cassel:2009wt,Slatyer:2009vg,Feng:2010zp,Kamada:2023iol}. Furthermore, when the mediator is sufficiently light or the relative velocity of the annihilating particles is sufficiently small, dark matter bound-state formation can also become important, as it is governed by the same long-range interaction effects~\cite{Petraki:2016cnz,Biondini:2023ksj}.
In this work, we investigate the enhancement of the annihilation rate of a pair of slow-moving finite-size dark matter particles arising from interactions mediated by the exchange of dark $\rho$ or $\pi$ mesons, which we refer to as the finite-size Sommerfeld effect. In our previous study, we found that the introduction of a finite-size parameter suppresses the Sommerfeld enhancement and that the parameters corresponding to the peaks of the enhancement factor are closely related to those characterizing different regimes of the self-scattering cross section of puffy dark matter particles~\cite{Wang:2023wbw,Tulin:2017ara,Colquhoun:2020adl}. In this work, we revisit the impact of finite-size effects on the $s$-wave Sommerfeld enhancement in a more systematic and detailed manner. By introducing two dimensionless parameters, we characterize the resonance structure of the Sommerfeld enhancement factor in the presence of an additional size parameter. We find that, in contrast to the point-particle case, the inclusion of finite-size effects causes the zero-energy resonance peak to broaden into a finite-width region in parameter space. We argue that this is a natural physical consequence of the finite spatial extent of the dark matter particle and therefor provides further support for the appropriateness of our choice of dimensionless parameters. Furthermore, we expect that a similar finite-width zero-energy resonance structure may also appear in the self-scattering of finite-size dark matter particles, since both phenomena are governed by solutions of the Schrödinger equation with the same interaction potential. This expectation provides an additional motivation for reexamining the finite-size Sommerfeld effect and its phenomenological implications.

This paper is organized as follows. In Sec.\ref{sec2}, we consider the interaction between finite-size dark matter particles without assuming any specific internal structure. The corresponding Schrödinger equation in the $s$-wave channel is solved numerically. A detailed study of the finite-size Sommerfeld effect is then carried out. We further analyze the impact of the size parameter on the zero-energy resonance behavior of the Sommerfeld enhancement factor. In Sec.\ref{sec3}, we consider a nugget-type dark matter model and analyze the resonance behavior of the Sommerfeld enhancement using the constraints associated with its specific internal structure. Sec. \ref{sec4} gives our conclusions.

\section{Finite-size Sommerfeld enhancement effect}\label{sec2}

In this section, we study the Sommerfeld enhancement for finite-size dark matter in the $s$-wave. While the enhancement for point-like dark matter is known to arise from non-perturbative interactions in the low-velocity limit, the finite size of dark matter $R_{\chi}$ introduces an additional length scale that may substantially affect the enhancement mechanism. In our previous work, we investigated the relation between self-scattering and self-annihilation for puffy dark matter and showed that finite-size effects can lead to a suppression of the Sommerfeld enhancement~\cite{Wang:2023wbw}. Here, we revisit this issue using a different numerical approach and perform a detailed comparison between the finite-size and point-like cases~\cite{Iengo:2009ni}. Particular attention will be paid to the resonance behavior of the enhancement factor, which has not been systematically explored in the context of finite-size dark matter. Before presenting our new analysis, we will also briefly recapitulate the formalism and numerical framework used to incorporate finite-size effects into the calculation of the Sommerfeld factor.

In the non-relativistic limit, the full annihilation cross section of two puffy dark matter particles can be expressed as the product of the bare annihilation cross section $\sigma_{0,l}$ and the Sommerfeld enhancement factor $S$, namely, $\sigma_l=\sigma_{0,l}\times S$~\cite{Arkani-Hamed:2008hhe}. For the $s$-wave  ($l=0$), the Sommerfeld enhancement factor can be obtained by solving the radial Schrödinger equation for the reduced two-body wave function $R_{p,l}$. The corresponding Schrödinger equation is given by 
\be \label{eq1}
\left(\frac{d^{2}R_{p,l}}{dr^{2}}+\frac{2}{r}\frac{dR_{p,l}}{dr}-\frac{l(l+1)R_{p,l}}{r^{2}}\right)+\left(p^{2}-2\mu V(r)\right)R_{p,l}=0,
\ee
where we define the center-of-mass (CM) momentum of the incoming particles $p=\mu v$ and the reduced dark matter mass $\mu=M_{\chi}/2$.  For the puffy dark matter the potential $V(r)$~\footnote{Note that finite-size dark matter is also called puffy dark matter. The puffy dark matter framework was first introduced in Ref.~\cite{Chu:2018faw}, where it was shown that the interaction Hamiltonian between two puffy dark matter particles can be constructed solely from their dark charge distributions in momentum space. The corresponding interaction potential in coordinate space was subsequently derived rigorously in Ref.~\cite{Wang:2023xii}. In the present work, we adopt this framework and neglect the strong interactions among the dark constituent particles.} is given by  
\begin{align}\label{eq2}
V(r) & = \begin{cases}
~- g(r,y) & r<2R_{\chi}\, , \\
\hspace{5cm}\  & \ \\[-6.mm]
~-\alpha\frac{e^{-m_{\phi}r}}{r} 
\times h\left(y\right)&  r>2R_{\chi}\,,
\end{cases}
\end{align}
where $y=R_{\chi}m_{\phi}$ is  the ratio between the radius of a puffy dark matter particle and the force range among these two particles, $r$ is the relative distance between dark matter particles, and the dark fine structure constant $\alpha = g_{\chi}^2/4\pi$ with $g_{\chi}$ being the coupling constant. The specific form of $V(r)$ is presented in Appendix~\ref{appa} . The detailed deduction of this potential  can be found in the appendix  of  Ref.{\cite{Wang:2023xii}}.

\begin{figure}[ht]
\centering
\includegraphics[width=8.5cm]{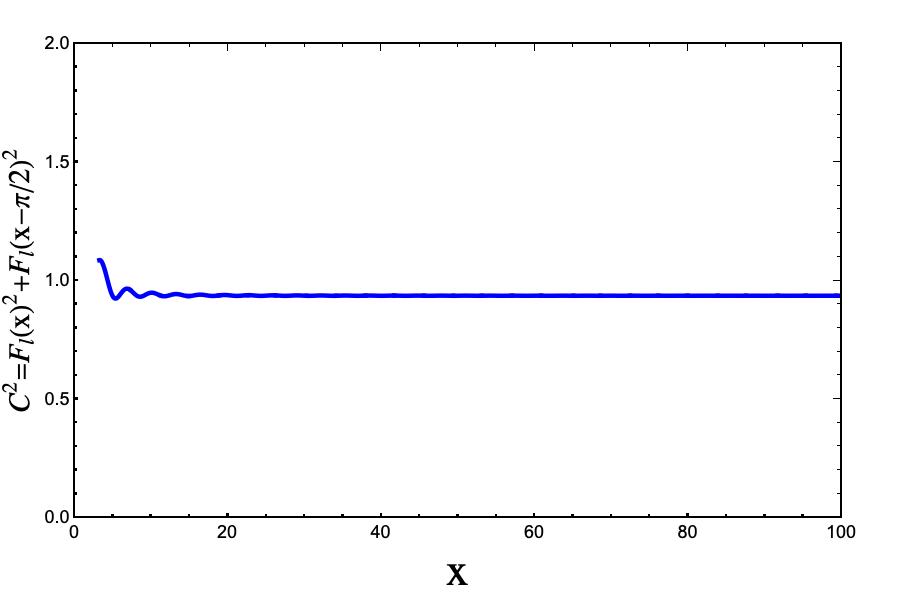}		
\caption{The plot of $F_0(x)^2+F_0(x-\pi/2)^2$ versus $x$ for $m_{\phi}=1~\rm GeV$, $R_{\chi}=1~\rm GeV$, $m_{\chi}=10~\rm GeV$, $\alpha=0.01$ and $v/c=0.01$.}
\label{fig1-1}
\end{figure}

By introducing the dimensionless variable $x=pr$ and writing the radial wave function as $R_{p,l}=Npx^l\phi_l(x)$, the radial Schrödinger equation can be recast into the form
\be \label{eq3}
\phi''-\frac{\phi_{l}(x)}{x^{2}}l(l+1)+\left(1-\frac{2}{ap}V(\frac{x}{p})\right)\phi_{l}(x)=0,
\ee
where $a=v/\alpha$. Following Ref.~\cite{Iengo:2009ni}, the initial condition is set as $\phi(x)_{x\rightarrow0}\rightarrow x^{l+1}$. The asymptotic behavior of  the wave function  will be 
\be \label{eq4}
\phi_l(x)_{x\rightarrow\infty}\rightarrow C\sin(x-\frac{l\pi}{2}+\delta_l),
\ee
where $\delta_l$ is the phase shift with $l$ wave. And using a constant $C$, the Sommerfeld enhancement factor is obtained as 
\be \label{eq5}
S=\left( \frac{1\cdot 3\cdots (2l+1)}{C} \right)^2.
\ee
In order to get the precise solution of $C$, we define a function $F_l(x)=\phi(x)$ and we can plot $F_l(x)^2+F_l(x-\pi/2)^2$ for a large $x$. When it is a constant line, its value is equal to $C^2$ (Note that this works well for $l=0$ as shown in Fig.~\ref{fig1-1}. For $l\neq0$ case, the improved approach can be found in Ref.\cite{Iengo:2009ni}).

\begin{figure}[ht]
	\centering
	\includegraphics[width=5.05cm]{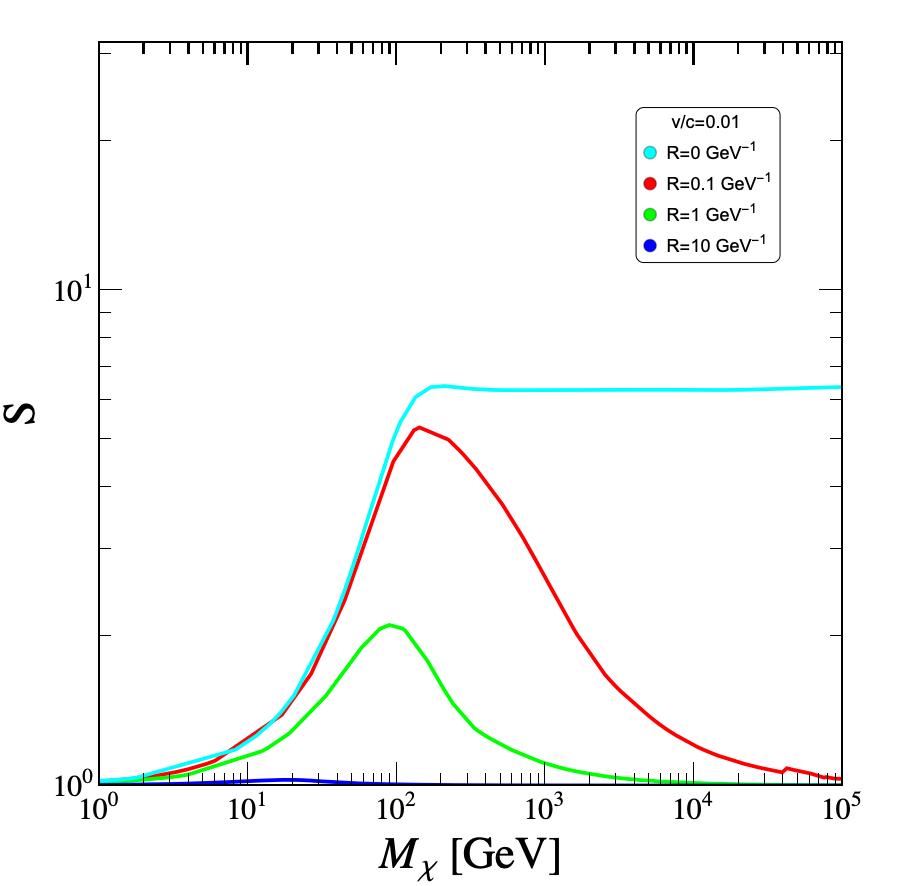}		
	\includegraphics[width=5.05cm]{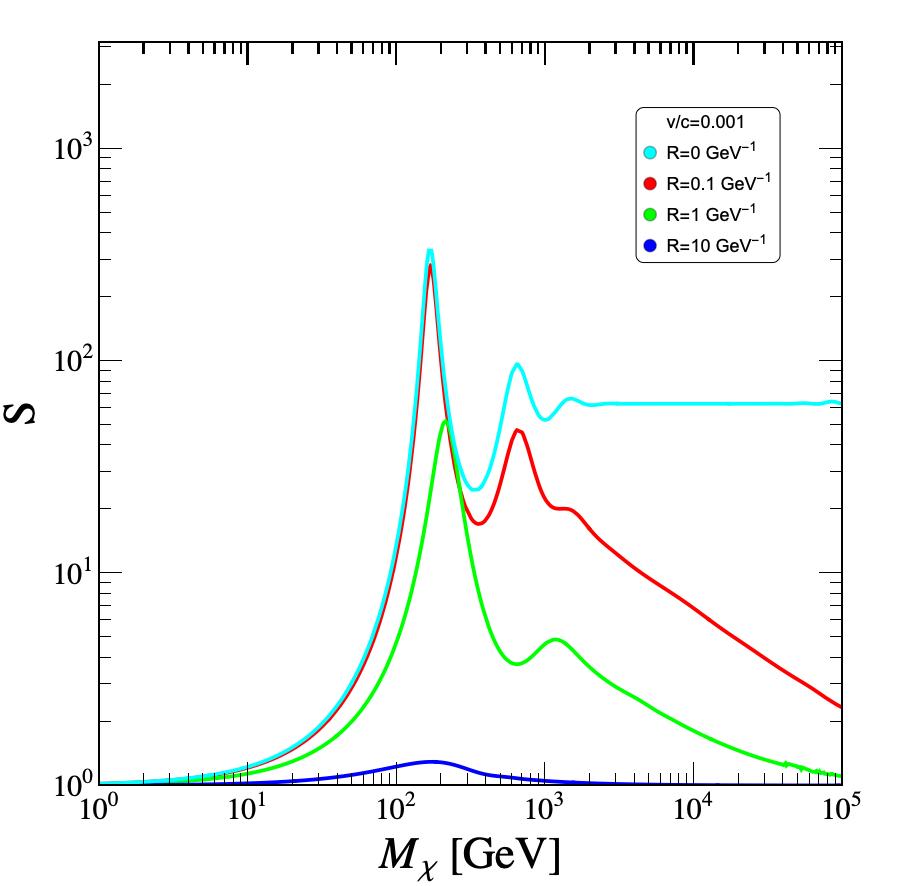}		
	\includegraphics[width=5.05cm]{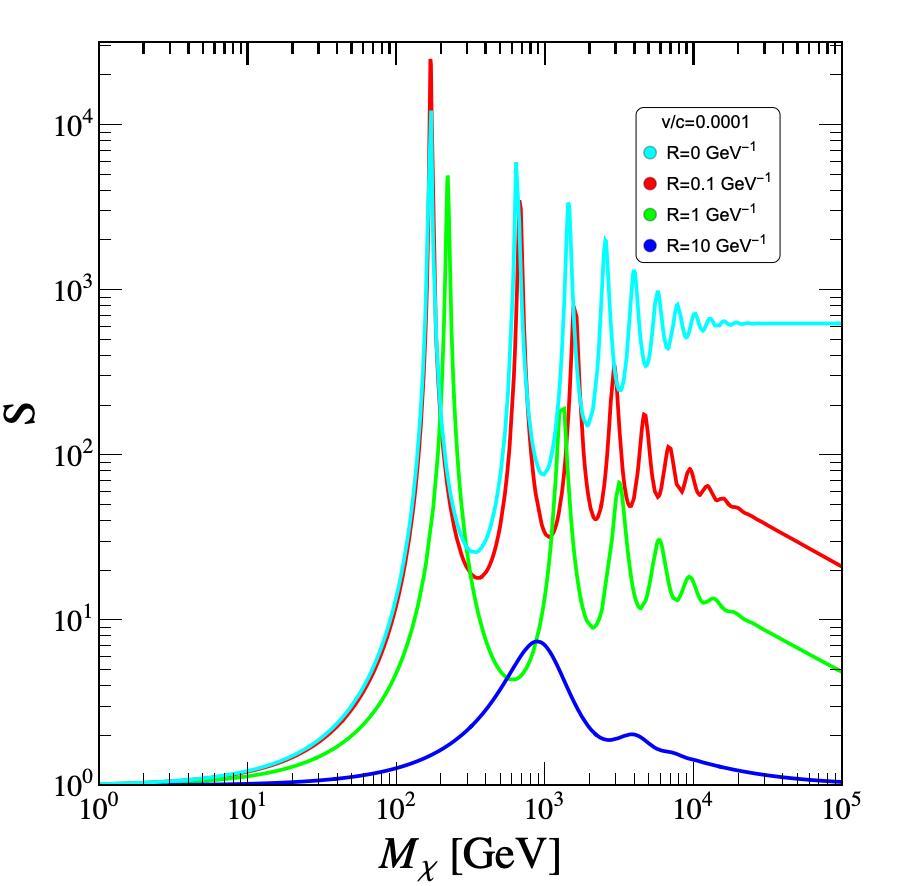}\\		
	\caption{Sommerfeld enhancement factor for the Yukawa and puffy potentials with $\alpha=0.01$ at different velocities: left panel ($v/c=0.01$), middle panel ($v/c=0.001$), and right panel ($v/c=0.0001$).
} 
	\label{fig2}
\end{figure}

We now turn to our new analysis, i.e., the impact of the size parameter on the Sommerfeld enhancement factor. By substituting the Yukawa potential and the puffy potential into Eq.~(\ref{eq3}), respectively, we obtain the Sommerfeld enhancement factor for different dark matter radii, as shown in Fig.~\ref{fig2}. From Fig.~\ref{fig2} we see that the peak value of the Sommerfeld enhancement factor decreases as the dark matter radius increases. This suppression becomes particularly pronounced for a large dark matter mass. Specifically, when the dark matter mass exceeds $10^4~\rm GeV$, the enhancement factor drops rapidly in the finite-size case, whereas it approaches an approximately constant value for point-like dark matter. This behavior can be understood from the relative importance of the kinetic and potential terms in the Schrödinger equation. For point-like dark matter, increasing the dark matter mass effectively suppresses the Yukawa potential (note that the potential function here depends on $x/p$ as its independent variable), causing the kinetic term to dominate the dynamics as in  Eq.~(\ref{eq3}). In contrast, for the puffy potential, the effect of increasing the dark matter mass cannot be considered independently, since the magnitude of the potential also depends on other parameters, such as the dark matter radius and velocity. As a result, the competition between the kinetic and potential terms remains nontrivial. Consequently, Fig.~\ref{fig2} mainly exhibits the modification of the resonance structure of the Sommerfeld enhancement factor as the dark matter mass varies. In particular, the locations of the resonance peaks are shifted relative to those in the point-particle case. Furthermore, a comparison between the blue curve and the other curves in Fig.~\ref{fig2} shows that the resonance width becomes narrower as the dark matter radius increases. Therefore, the Sommerfeld effect for puffy dark matter exhibits a much richer and more complicated resonance structure than that of point-like dark matter due to the additional dynamical scale introduced by the finite size. It is consequently difficult to identify a simple universal pattern governing the resonance behavior. These results further demonstrate that finite-size effects constitute an additional factor, beyond the low-velocity condition, that can significantly influence the non-perturbative dynamics responsible for Sommerfeld enhancement. These observations indicate that the finite size of dark matter introduces qualitatively new features into the non-perturbative dynamics of the annihilation process. The size effect therefore represents an additional control parameter of the Sommerfeld enhancement, alongside the conventional velocity dependence, and may play a dominant role in determining the resonance structure in certain regions of parameter space.

\begin{figure}[ht]
	\centering
	\includegraphics[width=5.05cm]{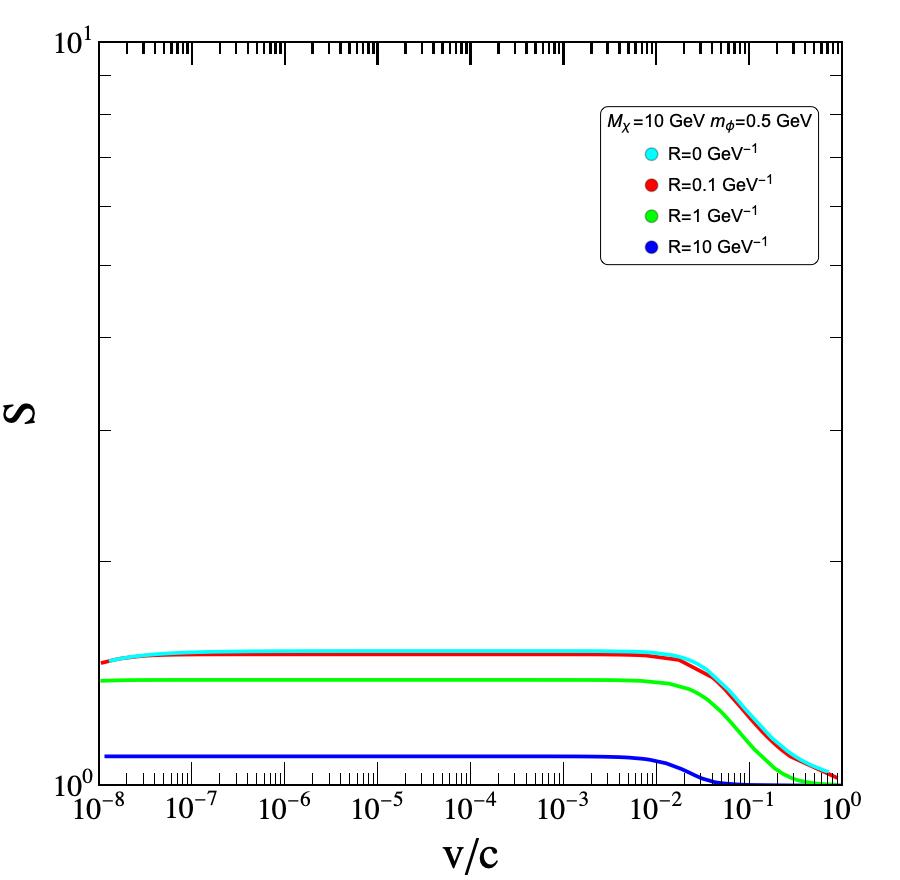}		
	\includegraphics[width=5.05cm]{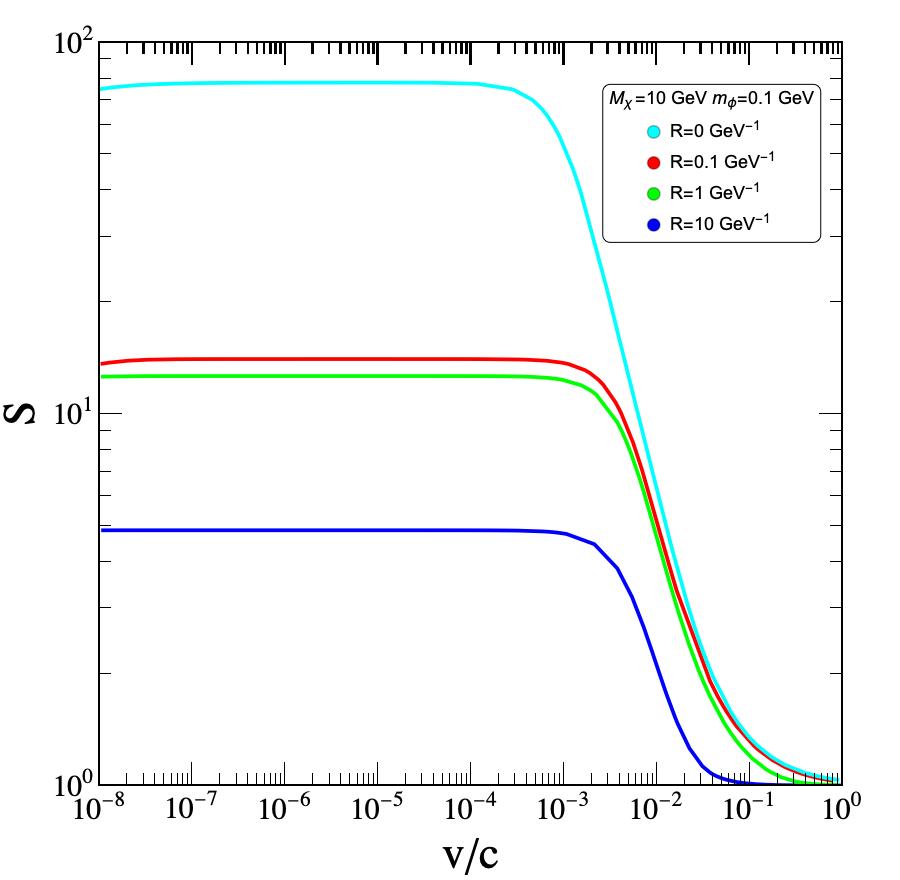}		
	\includegraphics[width=5.05cm]{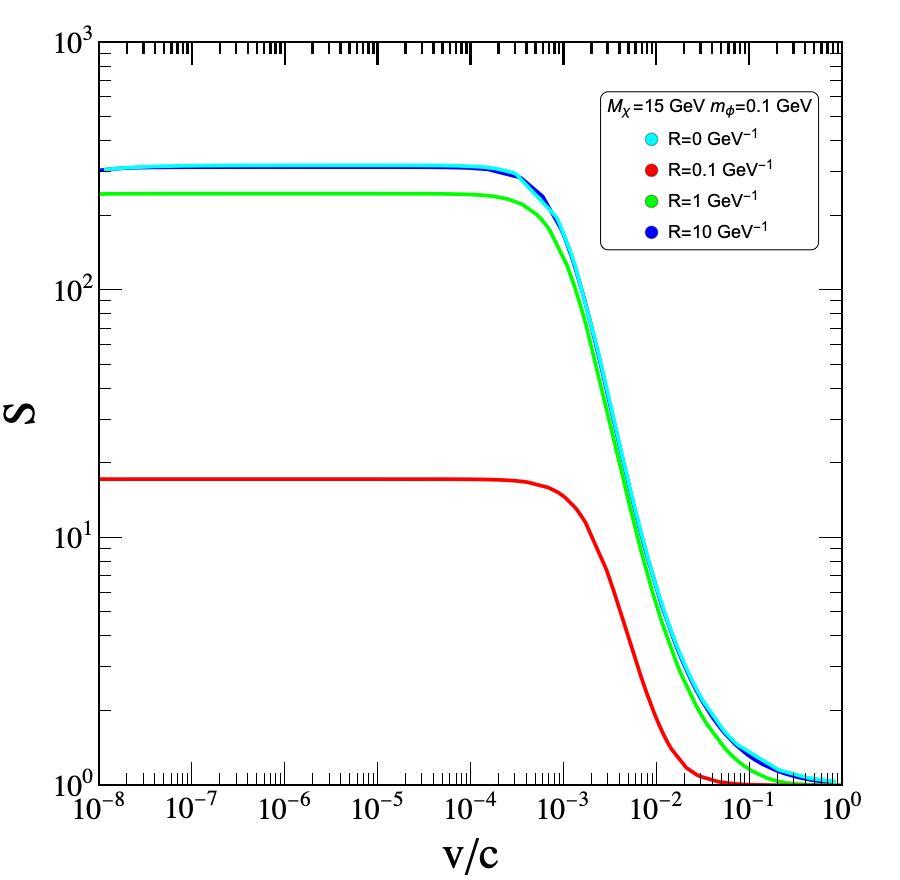}\\		
	\caption{Velocity- dependence of the Sommerfeld enhancement factor for different dark matter radii, dark matter masses $M_{\chi}$, and mediator masses $m_{\phi}$, with $\alpha=0.01$.} 
	\label{fig3}
\end{figure}

We further investigate the velocity dependence of the Sommerfeld enhancement in the presence of finite-size effects. As shown in Fig.~\ref{fig3}, both the left and middle panels demonstrate that the Sommerfeld enhancement is generally suppressed as the dark matter radius increases. This behavior is consistent with the results discussed above and confirms that finite-size effects tend to reduce the magnitude of the enhancement factor. In particular, the comparison between the red curve $R_{\chi}=0.1~\rm GeV^{-1}$ and the cyan curve corresponding to the point-particle case in the middle panel shows that even a relatively small dark matter radius can lead to a substantial modification of the Sommerfeld enhancement. This indicates that finite-size effects cannot always be neglected, even when the characteristic size of the dark matter particle is small. Interestingly, the right panel of Fig.~\ref{fig3} shows that the enhancement factors for $R_{\chi}=1~\rm GeV^{-1}$ and $R_{\chi}=10~\rm GeV^{-1}$ are quite close to that of the point-like dark matter. This indicates that the system approaches a resonant regime in which the finite-size suppression is compensated by resonance effects, resulting in a Sommerfeld enhancement comparable to the point-particle limit. Taken together, the results of Fig.~\ref{fig3} demonstrate that the finite-size effects play a crucial role in determining the velocity dependence of the Sommerfeld enhancement. Besides the well-known low-velocity enhancement, the dark matter size introduces an additional scale that can significantly alter the resonance structure and the magnitude of the enhancement factor.

\begin{figure}[ht]
	\centering
	\includegraphics[width=6.9cm]{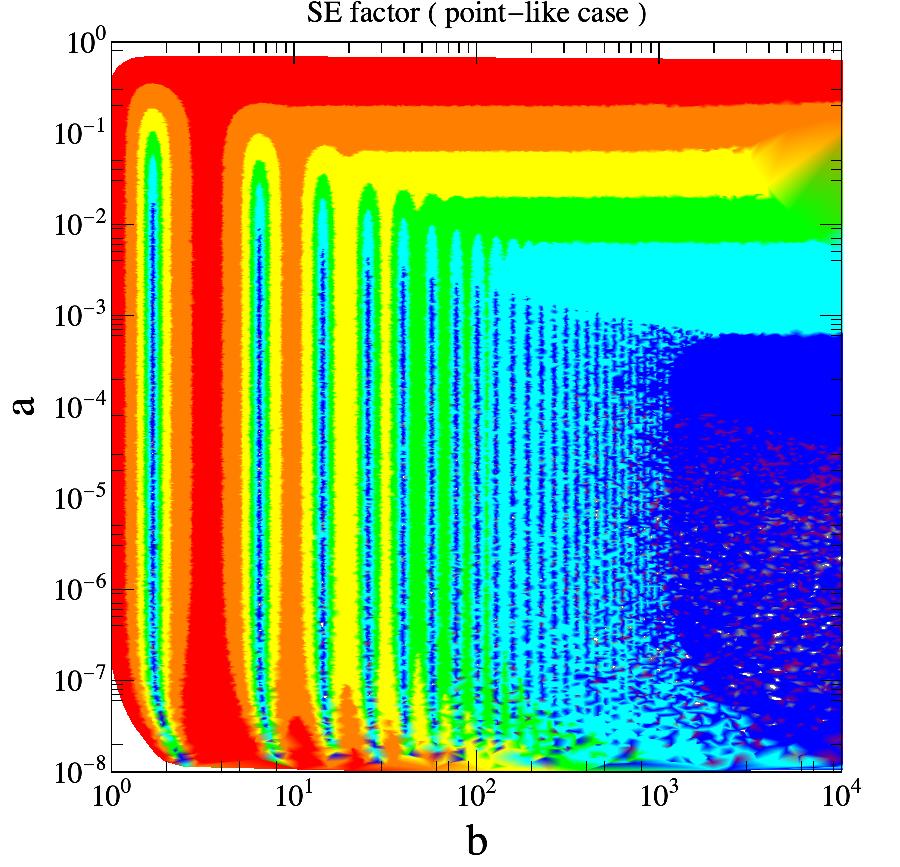}		
	\includegraphics[width=8.05cm]{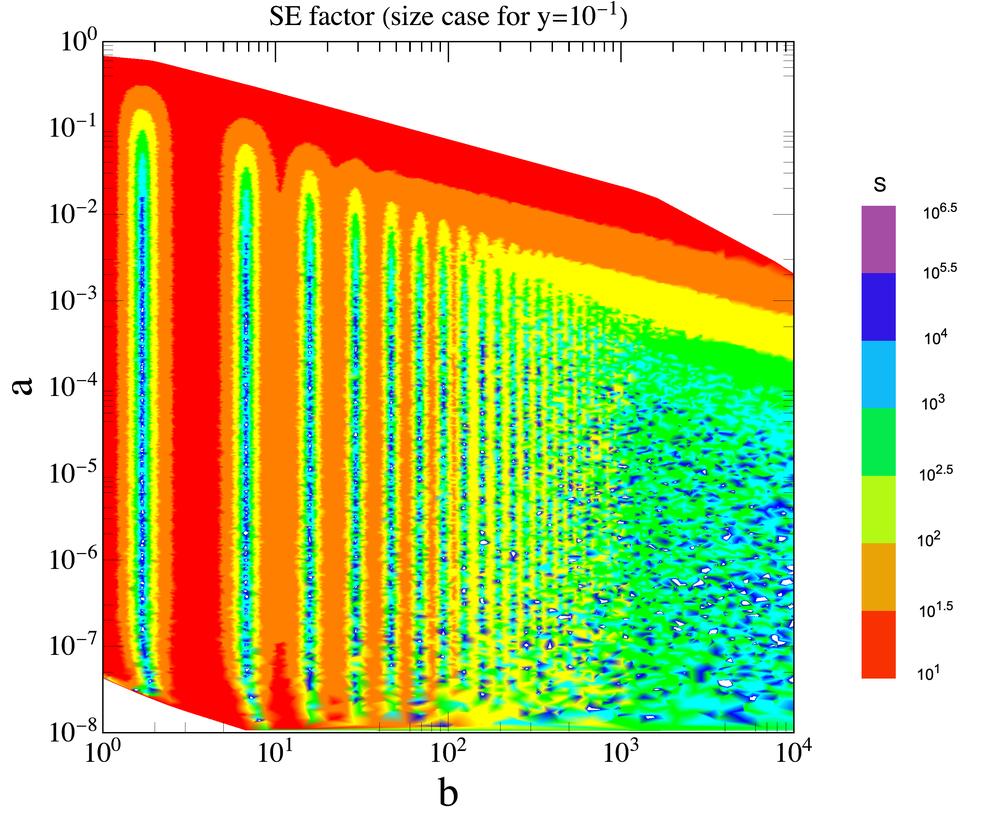}\\	
	\includegraphics[width=6.9cm]{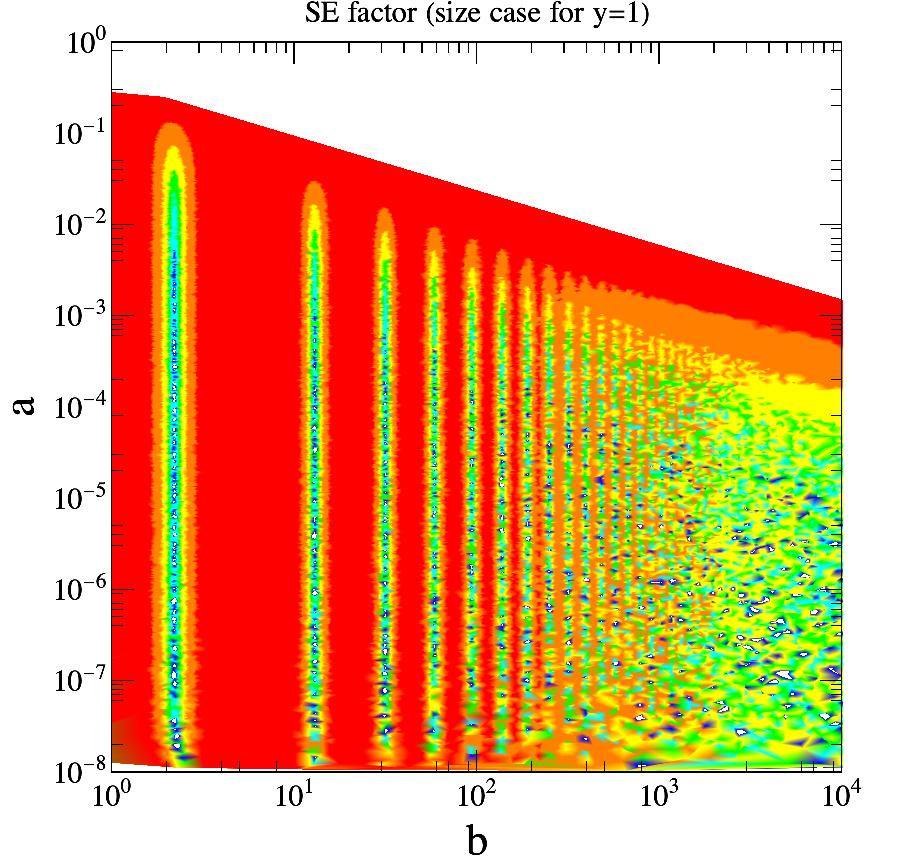}			
	\includegraphics[width=8.05cm]{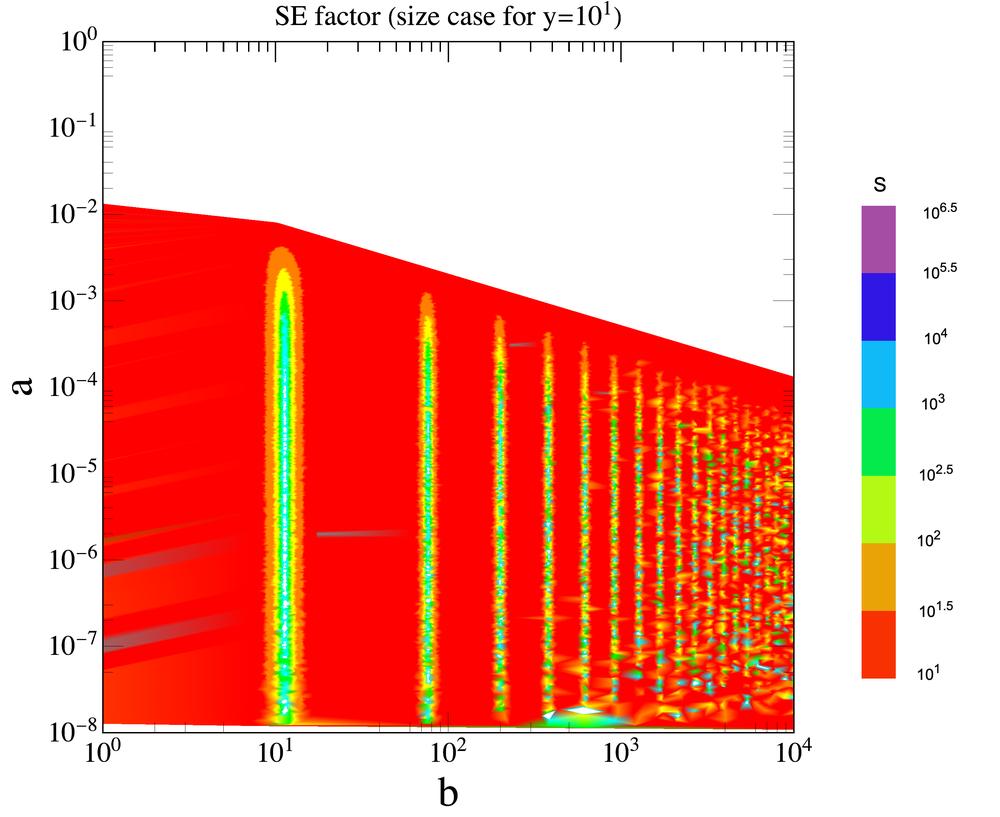}\\	
	\caption{Sommerfeld enhancement factor in the plane of $a$ versus $b$ for point-like dark matter and for puffy dark matter with different values of $y$.
}
	\label{fig4}
\end{figure}

Thus, to systematically investigate the impact of the finite-size effects on the Sommerfeld enhancement, we parameterize the system in terms of two dimensionless quantities $a$ and $b=\alpha m_{\chi}/m_{\phi}$ as introduced in Ref.~\cite{Masi_2016}. The resulting Sommerfeld enhancement factors for different values of the radius-to-force-range ratio $y$ are presented in Fig.~\ref{fig4}. As can be seen from Fig.~\ref{fig4}, the resonant behavior appears for specific $b$ values. For puffy dark matter, after fixing a given radius-to-force-range ratio $y$, the resonance positions are found to shift toward larger values of $b$ as the radius-to-force-range ratio increases. In other words, the resonance peaks move to the right in parameter space with increasing finite-size effects. A comparison between the panels corresponding to finite-size dark matter (the lower panels and the upper-right panel of Fig.~\ref{fig4}) and the point-particle case (the upper-left panel) further reveals that the finite-size effects can eliminate the Sommerfeld enhancement in certain regions of parameter space. Consequently, portions of the parameter space that exhibit enhancement in the point-particle limit become devoid of Sommerfeld enhancement once the finite-size effects are taken into account. Moreover, the strength of the resonant enhancement decreases as the radius-to-force-range ratio $y$ increases. This trend is particularly evident from the first resonance peak shown in the lower-right panel of Fig.~\ref{fig4}, where the corresponding Sommerfeld enhancement factor is significantly smaller than those obtained for the other two radius-to-force-range ratios $y$. These results indicate that the finite size of dark matter modifies both the location and the strength of the Sommerfeld resonances. As the radius-to-force-range ratio $y$ increases, the resonance structure becomes progressively weaker and is displaced toward larger values of $b$, leading to a substantial reduction of the enhancement in parts of the parameter space.

The results presented above suggest that the resonance structure of the finite-size Sommerfeld enhancement may also be governed by a small number of dimensionless combinations of the underlying parameters. We therefore search for an analogue of the ($a$,$b$) parametrization shown in Fig.~\ref{fig4}, with the aim of identifying the resonance locations in a universal manner. A natural starting point is provided by the self-scattering of puffy dark matter. As demonstrated in Ref.~\cite{Wang:2023wbw}, the self-scattering cross section can be organized in terms of the two dimensionless parameters $bf(y)$ and $\sqrt{y}a/f(y)$. In particular, the Born regime is characterized by the condition $bf(y)\ll 1$. Since both self-scattering and Sommerfeld enhancement originate from the same underlying Schrödinger dynamics, it is plausible that these parameters may also play a central role in determining the resonance structure of the finite-size Sommerfeld enhancement. For later convenience, we reproduce below the Born-limit expression obtained in Ref.~\cite{Wang:2023wbw}:
\small
\begin{eqnarray}
m_\chi 
\left|\int_{0}^{\infty}rV(r)dr\right|
= \frac{ \alpha m_{\chi}}{m_{\phi}}
\frac{3(-15+y^2(15-10y+4y^3)+15(1+y^2)e^{-2y})}{10y^6}=bf(y) \ll 1\,,
\label{pufb}
\end{eqnarray}
\normalsize

where  function $f(y)$ is defined for the estimation of the validity. 

\begin{figure}[ht]
	\centering
	\includegraphics[width=8.5cm]{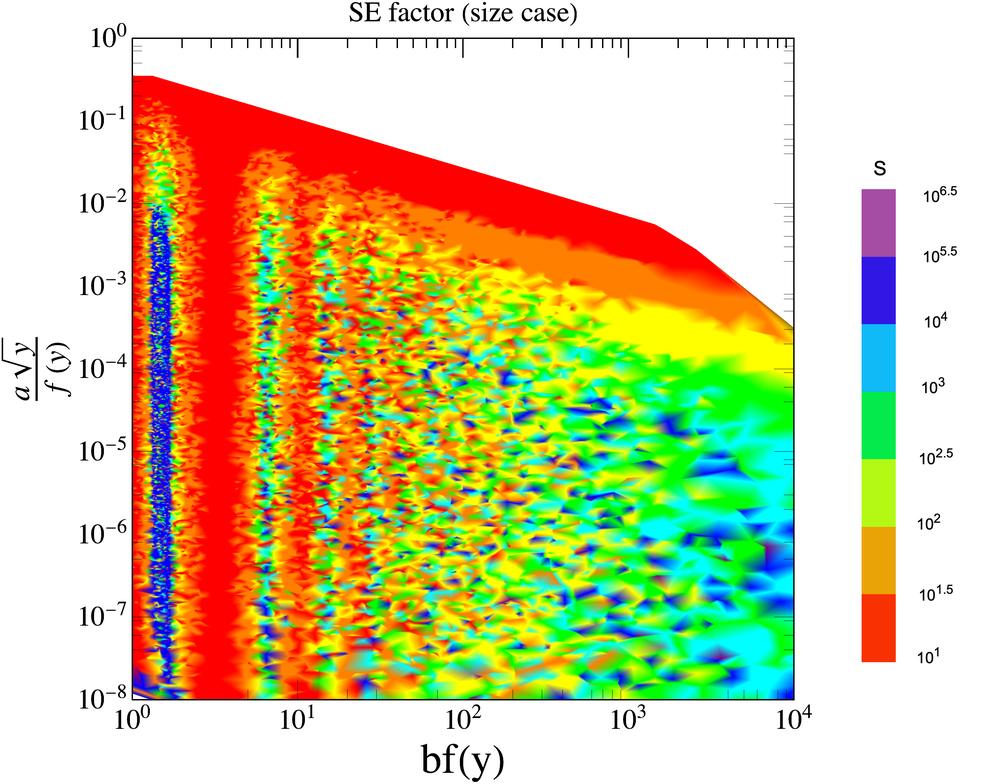}		
	\includegraphics[width=6.8cm]{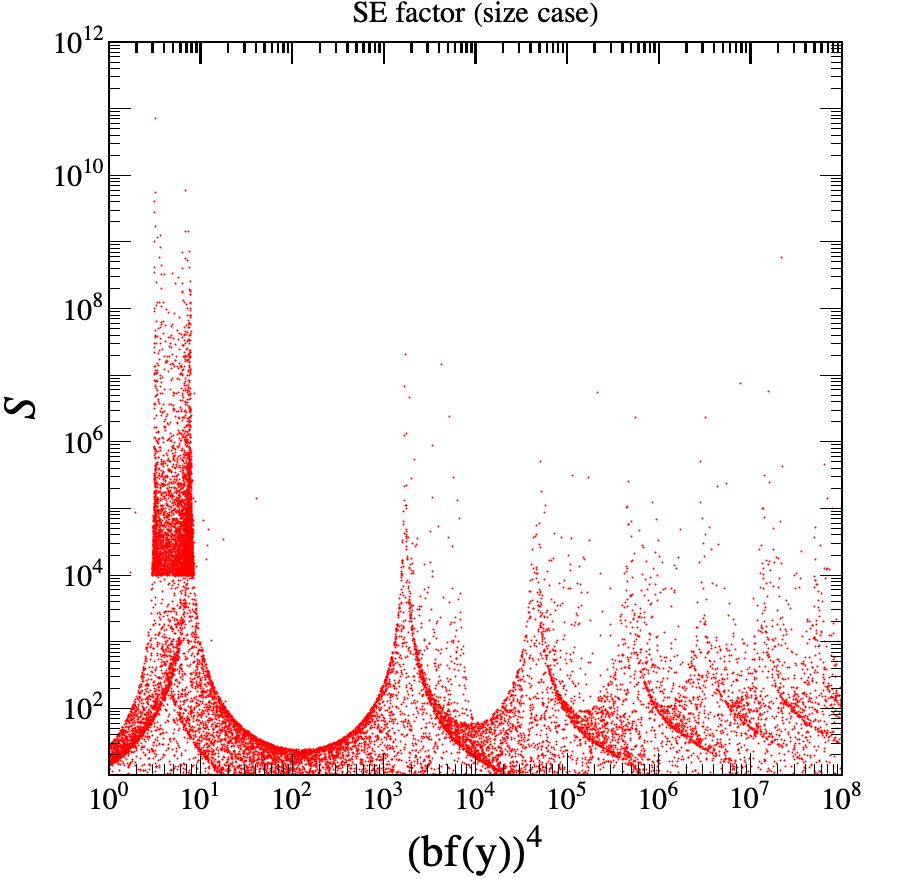}\\	
	\caption{Left panel: Sommerfeld enhancement factor in the $\sqrt{y}a/f(y)$ versus $b f(y)$ plane for puffy dark matter annihilation; Right panel: Sommerfeld enhancement factor $S$ as a function of $(b f(y))^4$ where the fourth power is chosen to make the width of the resonance more visible.
}
	\label{fig5}
\end{figure} 

As can be seen from the classification of the self-scattering cross section for puffy dark matter shown in the right panel of Fig.~2 of Ref.~\cite{Wang:2023wbw}, resonant behavior can indeed be characterized by the two dimensionless parameters $bf(y)$ and $\sqrt{y}a/f(y)$. However, the boundaries between different resonant regions are not sharply defined. Since the non-perturbative effects responsible for both self-scattering and self-annihilation originate from the solution of the same Schrödinger equation, it is natural to investigate whether these two dimensionless parameters can provide a clearer characterization of the resonance boundaries in the Sommerfeld enhancement associated with dark matter annihilation.  The corresponding results are presented in the left panel of Fig.~\ref{fig5}. We find that the resonant states can indeed be clearly distinguished in terms of these two dimensionless parameters, with well-separated resonance bands. Nevertheless, the resonance peaks are not sufficiently sharp and instead exhibit a finite width. To make this feature more apparent, we enlarge the horizontal axis by plotting the enhancement factor as a function of $(bf(y))^4$. As shown in the right panel of Fig.~\ref{fig5}, the first resonance peak clearly possesses a nonzero width. This observation provides a natural explanation for the blurred resonance boundaries observed in the self-scattering cross section: the resonance condition is realized over a finite interval of parameter space rather than at a single critical value.  
\begin{figure}[ht]
	\centering
	\includegraphics[width=7.65cm]{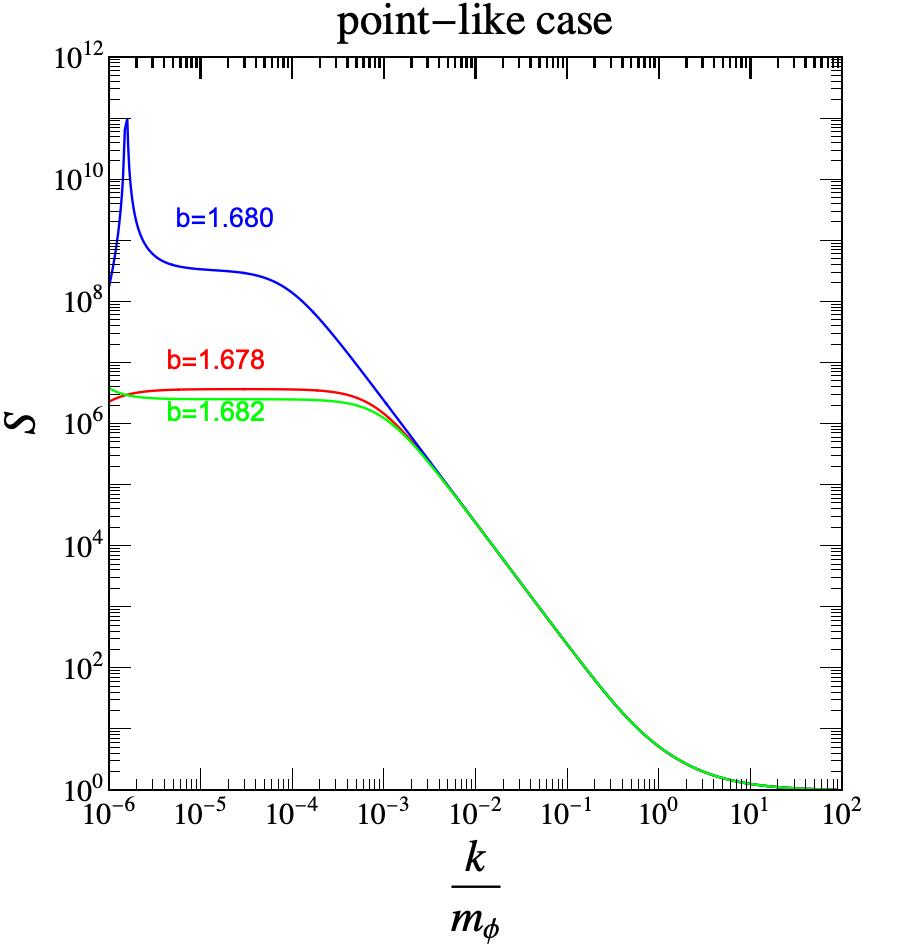}		
	\includegraphics[width=7.5cm]{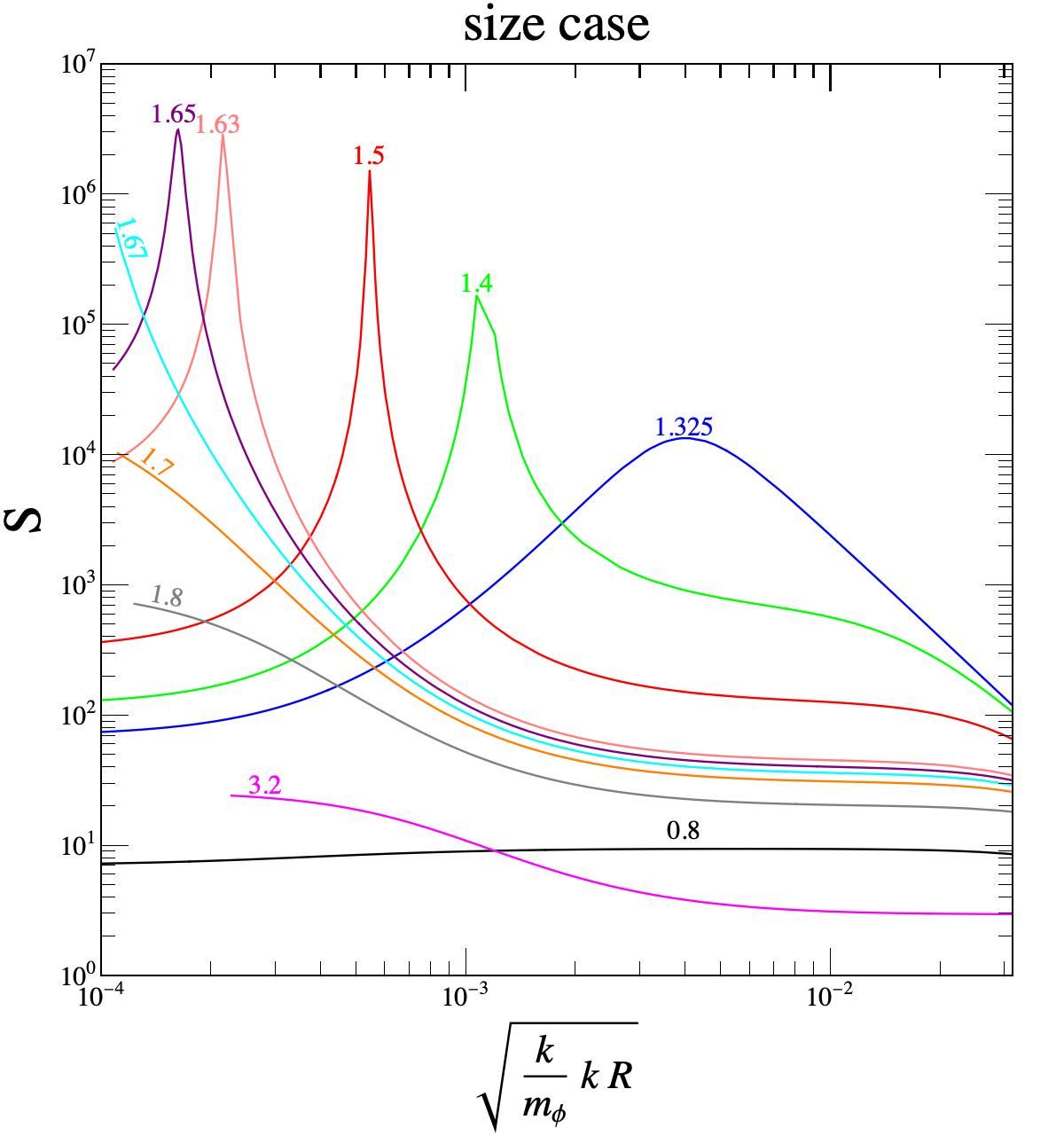}\\	
	\caption{Left panel: Sommerfeld enhancement factor $S$ as a function of $k/m_{\phi}$ with different values of parameter $b$ (
    near and at the first zero-energy resonance) for the point-like case;
    Right panel: Sommerfeld enhancement factor $S$ as a function of $\sqrt{(k/m_{\phi})(k R_{\chi})}$ with different values of parameter $bf(y)$ (near and at the first zero-energy resonance) for puffy dark matter.
   }
	\label{fig6}
\end{figure} 

To further verify this interpretation, we follow the analysis of Ref.~\cite{Kamada:2023iol}. We first reproduce the location of the first resonance in the point-particle case. The result is shown in the left panel of Fig.~\ref{fig6}, where the condition $k/m_{\phi}>1$ serves as the criterion for distinguishing the quantum and classical regimes.  For puffy dark matter, we introduce the analogous dimensionless parameter $\sqrt{(k/m_{\phi})(k R_{\chi})}$ as the corresponding diagnostic quantity. We then evaluate the Sommerfeld enhancement for several representative values of $bf(y)$, shown by the solid curves of different colors in the right panel of Fig.~\ref{fig6}.  As can be seen, the first zero-energy resonance is not localized at a single value of $bf(y)$, but instead extends over a finite interval, approximately from $bf(y)=1.325$ to $bf(y)=1.65$. In contrast, the cases $bf(y)=0.8$ and $bf(y)=3.2$ lie below and above the first zero-energy resonance, respectively. These results provide direct evidence that finite-size effects constitute an additional factor controlling the Sommerfeld enhancement of puffy dark matter. The emergence of a finite resonance width can be traced back to the introduction of the size scale $R_{\chi}$. Without specifying the detailed internal structure of dark matter particle, the resonance states of the $s$-wave Sommerfeld enhancement are therefore expected to occupy a finite region of parameter space rather than a single resonance point. In other words, finite-size effects transform the conventional resonance condition into a resonance band.  For dark matter candidates with a specific internal structure, such as nugget-type dark matter, the resonance behavior of the Sommerfeld enhancement can exhibit additional features, which will be discussed in the next section.

\section{The results for nugget-type dark matter}\label{sec3}

As in the visible sector, where matter is composed of finite-size particles such as protons, neutrons, mesons, and atoms, there exist many finite-size dark matter candidates with well-defined internal structures, including dark atoms, dark glueballs, and dark mesons. Once the internal structure of dark matter particle is specified, nontrivial relations among its mass, radius, self-coupling strength, and even the mediator mass are generally imposed. In this work, we focus on nugget-type dark matter. In our previous study of the direct detection of puffy dark matter, we found that for sufficiently large dark matter radii, finite-size effects drive the dark matter-nucleus scattering cross section into the Born regime, where non-perturbative effects associated with the size parameter become negligible~\cite{Xu:2025xaq}. A similar behavior is observed in the case of finite-size Sommerfeld enhancement. As shown in Figs.\ref{fig2}-\ref{fig4}, the Sommerfeld enhancement factor approaches unity for sufficiently large dark matter radii. Therefore, when investigating the Sommerfeld enhancement for nugget-type dark matter, we restrict our attention to systems containing a relatively small number of constituent particles, corresponding to small values of $N$. A systematic study of the properties of nugget-type dark matter can be found in Ref.~\cite{Gresham:2017zqi}. Here, we briefly summarize the relevant relations. The mass of a nugget can be written as $M_{\chi}=N m_{\chi}$, where $m_{\chi}$ denotes the mass of the constituent dark matter particle. Using the non-relativistic Fermi-gas approximation, the radius of the nugget can be estimated as $R_{\chi}\simeq
\left[81\pi^{2}/(4N g_{\rm dof}^{2}\alpha^{3}m_{\chi}^{3})
\right]^{1/3}$, where $g_{\rm dof}$ denotes the number of fermionic degrees of freedom. Details of this derivation can be found in Ref.~\cite{Wise:2014jva}. Once a specific internal structure is imposed on the puffy dark matter particle, the conditions for bound-state formation introduce additional constraints that can significantly affect the Sommerfeld enhancement. In particular, we are interested in how the resonance structure is modified when the enhancement factor is classified in terms of the two dimensionless parameters $bf(y)$ and $\sqrt{y}a/f(y)$. Questions of interest include the location of the first zero-energy resonance in terms of $bf(y)$, as well as the constraints on the dark matter mass arising from the internal structure and their implications for the Sommerfeld enhancement.

\begin{figure}[ht]
	\centering
	\includegraphics[width=8.3cm]{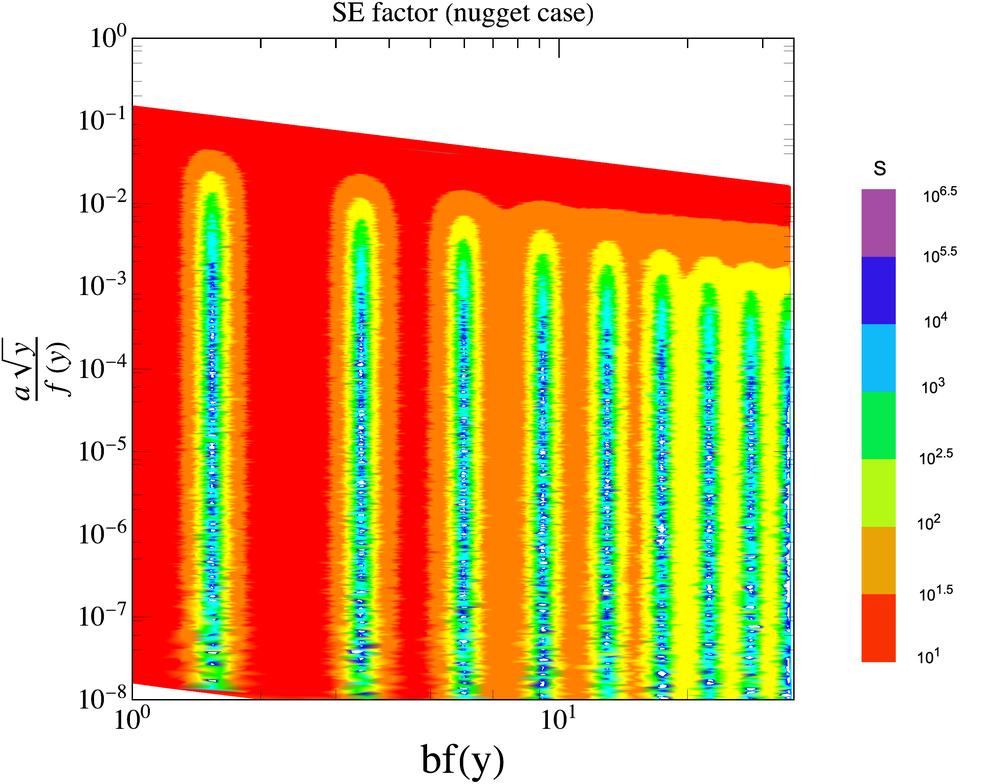}		
	\includegraphics[width=6.35cm]{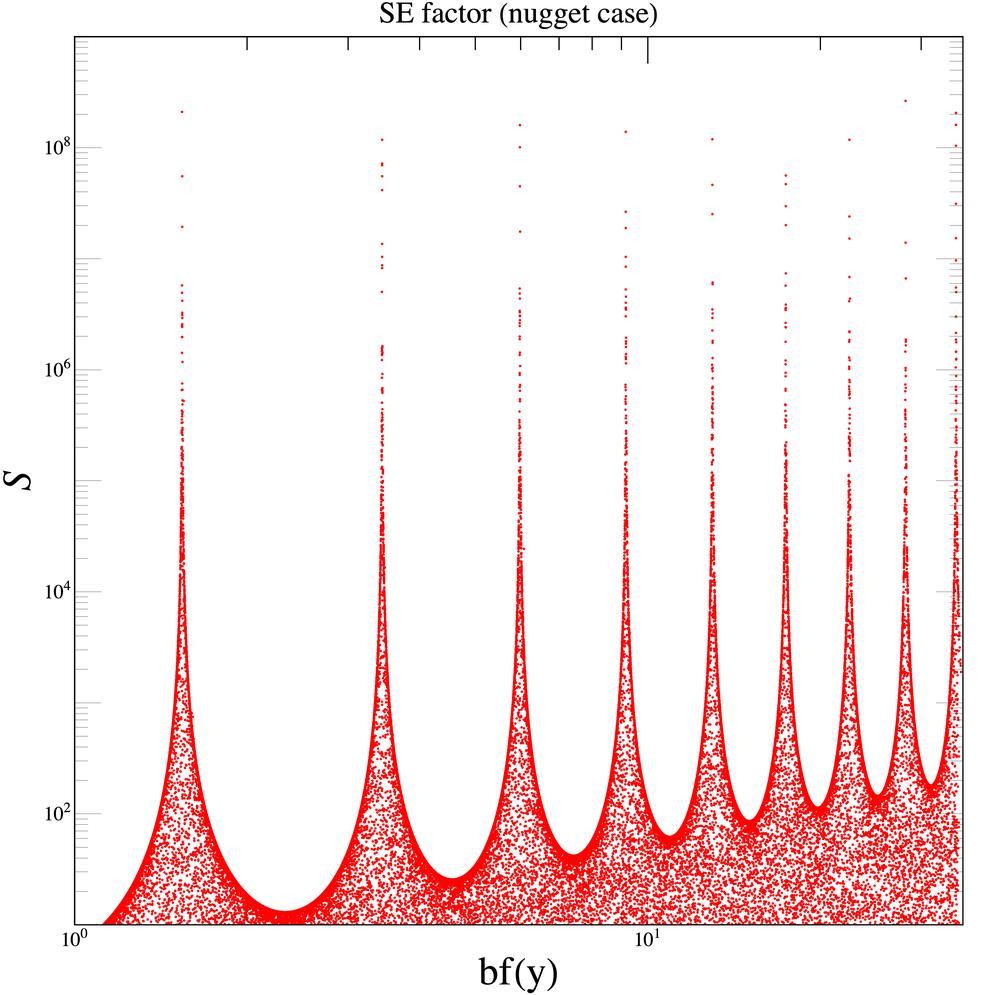}\\	
	\caption{Same as Fig.~\ref{fig5}, but the horizontal axis of the right panel is $b f(y)$ rather than $(b f(y))^4$.
}
	\label{fig7}
\end{figure} 
We now turn to the Sommerfeld enhancement of dark matter with a specific internal structure, namely nugget-type dark matter. The corresponding results are shown in Fig.~\ref{fig7}, where the number of constituent particles is fixed to $N=5$. The left panel displays the Sommerfeld enhancement factor in the parameter space spanned by the two dimensionless variables ($bf(y)$,$\sqrt{y}a/f(y)$). Comparing the left panel of Fig.~\ref{fig7} with Fig.~\ref{fig5}, one immediately notices that the finite width of the resonance peak disappears once a concrete internal structure is specified. In other words, the resonance band observed in the model-independent finite-size analysis collapses into a well-defined resonance line when the nugget structure is taken into account.  This feature is particularly evident in the right panel of Fig.~\ref{fig7}, where the resonance peak is plotted as a function of $bf(y)$. The location of the first zero-energy resonance is now associated with a definite value of $bf(y)$, analogous to the behavior found in the point-particle case. The key difference is that this resonance position depends on the number of constituent particles $N$. Consequently, the value of $bf(y)$ corresponding to the first zero-energy resonance may shift as $N$ varies. This result suggests that the resonance structure of the Sommerfeld enhancement may serve as a useful probe of the internal properties of puffy dark matter. In particular, the resonance position could potentially be used to distinguish between different dark matter internal structures and, in the case of nugget-type dark matter, to infer the number of constituent particles forming the bound state. 

\begin{figure}[ht]
	\centering
	\includegraphics[width=7.2cm]{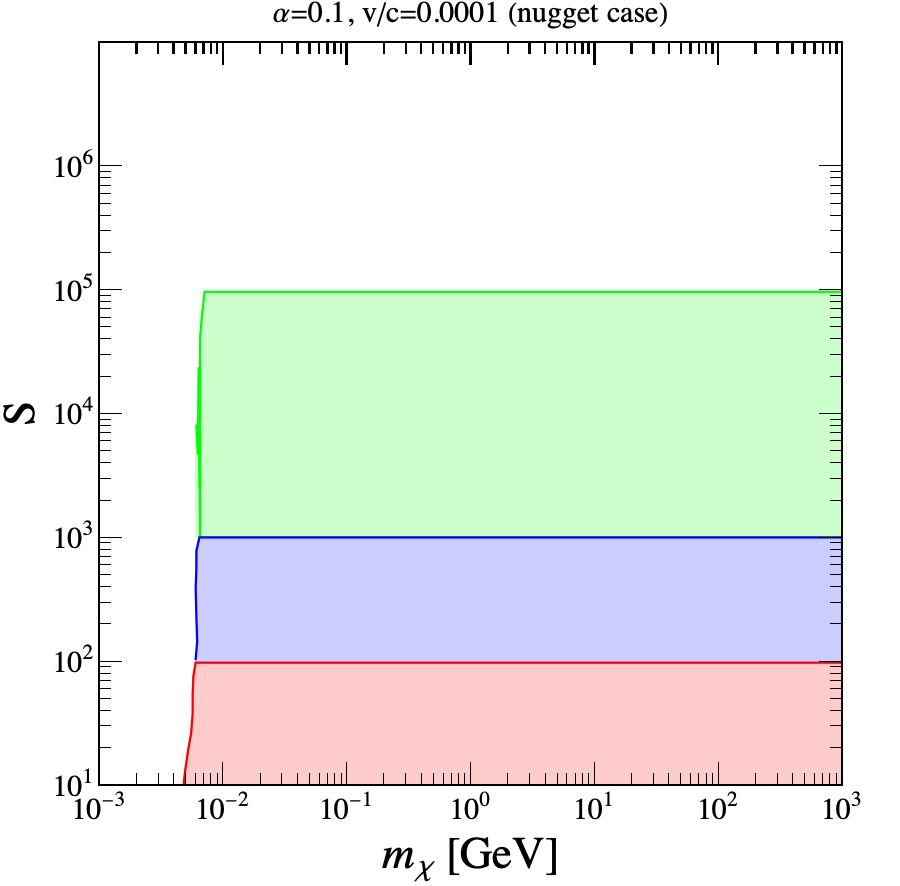}		
	\includegraphics[width=7.2cm]{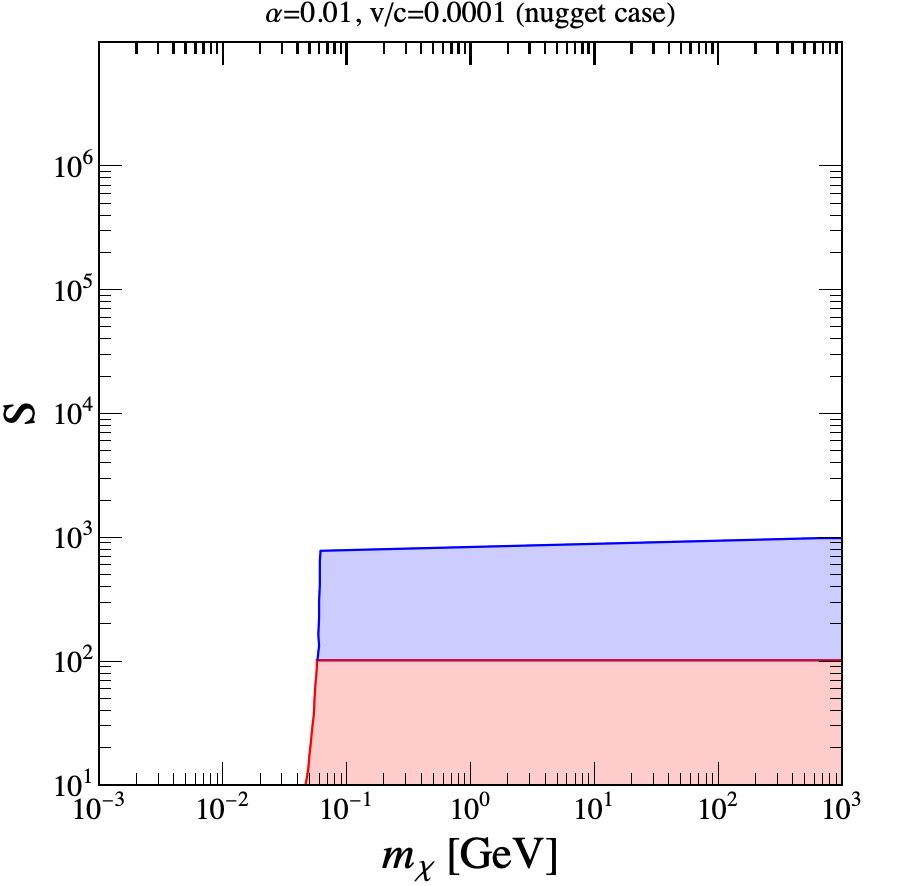}\\	
	\caption{Sommerfeld enhancement factor for nugget-type dark matter with $v/c=0.0001$ for different coupling constants: left panel ($\alpha=0.1$) and right panel ($\alpha=0.01$). Here the different color regions correspond to $S=10-100, 100-1000, 1000-10^5$, respectively. 
}
	\label{fig8}
\end{figure}

Finally, we fix the parameters of nugget-type dark matter at $N=5$ and $v/c=0.0001$, and scan over the dark matter mass and mediator mass. Fig.~\ref{fig8} shows the parameter space of  ($M_{\chi}$, $S$) for coupling constants  $\alpha=0.1$ (left panel) and $\alpha=0.01$ (right panel). The results indicate that fixing the internal structure of finite-size dark matter constrains the regions of Sommerfeld enhancement. Furthermore, as the value of $a=v/\alpha$ increases, i.e., in the right panel of Fig.~\ref{fig8}, the Sommerfeld enhancement for nugget-type dark matter weakens, leading to stronger constraints.

\section{Conclusion}\label{sec4}

If dark matter possesses a finite spatial extent, the non-perturbative Sommerfeld enhancement of dark matter annihilation is influenced not only by the low-velocity nature of the interaction but also by finite-size effects. Without specifying the internal structure of dark matter particle and considering only its charge distribution, we performed a detailed investigation of the impact of the size parameter on the Sommerfeld enhancement factor. We found that the finite-size effects modify the location of the zero-energy resonance and suppress the peak values of the Sommerfeld enhancement factor, with larger dark matter radii leading to stronger suppression. By introducing the two dimensionless parameters ($bf(y)$,$\sqrt{y}a/f(y)$), we showed that the resonance structure of the finite-size Sommerfeld enhancement can be systematically characterized. However, in the absence of a specified internal structure, the zero-energy resonance is no longer associated with a unique point in parameter space but instead exhibits a finite width. Once a concrete internal structure is imposed, the resonance behavior becomes analogous to that of point-like dark matter. In particular, the first zero-energy resonance is recovered at a well-defined parameter value rather than within a finite interval. This demonstrates that the resonance width observed in the model-independent finite-size analysis originates from treating the dark matter radius as an independent parameter. Our results therefore suggest that the position of the zero-energy resonance carries valuable information about the microscopic structure of puffy dark matter. In the future, measurements or indirect constraints on resonance locations may provide a means of discriminating different internal structures of puffy dark matter and offer new theoretical guidance for indirect dark matter searches. Consequently, the Sommerfeld enhancement may serve not only as a probe of dark matter interactions but also as a novel probe of the internal structure of puffy dark matter candidates.

\addcontentsline{toc}{section}{Acknowledgments}
\acknowledgments
 This work was supported by a Talent program from Chengdu Technological University (2024RC031), by the National Natural Science Foundation of China (NNSFC) under grant No. 12075300 and by a PI Research Fund from Henan Normal University (5101029470335). 

\appendix
\section{The  potential function}\label{appa}
The interaction potential between two finite particles is 

\begin{align}\label{eqa}
V(r) & = \begin{cases}
- g(y) & r<2R_{\chi} \\
\hspace{5cm}\  & \ \\[-6.mm]
-\alpha\frac{e^{-m_{\phi}r}}{r} \times h\left(y\right)&  r>2R_{\chi},
\end{cases}
\end{align}
where
\begin{eqnarray} \label{inbol}
g(r,y)&=&-\alpha4\pi^{2}(\frac{3}{4\pi R_{\chi}^{3}})^{2}\frac{(1+y)e^{-y}}{m_{\phi^{3}}}\frac{1}{2m_{\phi}^{3}} \nonumber \\
&& \times\left(\frac{2e^{-m_{\phi}(R_{\chi}+r)}(-1+e^{y})(e^{2y}(-1+y)+e^{y}(1+y))}{r}\right.\nonumber \\
&& \Bigg.+m_{\phi}(-e^{y}(2+m_{\phi}(r-2R_{\chi}))+e^{-y}(2-m_{\phi}r+2y))\Bigg)\nonumber \\
&& +\alpha4\pi^{2}\left(\frac{3}{4\pi R_{\chi}^{3}}\right)^{2}\frac{1}{m_{\phi}}\left(\frac{e^{-y}}{m_{\phi}^{2}}-\frac{e^{y}}{m_{\phi}^{2}}+\frac{R_{\chi}e^{-y}}{m_{\phi}}+\frac{R_{\chi}e^{y}}{m_{\phi}}\right) \nonumber \\
&& \times\frac{e^{-m_{\phi}(R_{\chi}+r)}(2+2y-e^{y}(2+m_{\phi}r(-2+m_{\phi}(r-2R_{\chi}))+2y))}{2m_{\phi}^{3}r}\nonumber  \\
&& +\alpha4\pi^{2}\left(\frac{3}{4\pi R_{\chi}^{3}}\right)^{2}\frac{1}{m_{\phi}^2}\left( \frac{1}{12}(r-2R_{\chi})(r+4R_{\chi})\right), \\
h\left(y\right)&=&4\pi\left(\frac{3}{4\pi R_{\chi}^{3}}\right)^{2}\frac{\pi}{m_{\phi}^{2}}\left(\frac{e^{-y}}{m_{\phi}^{2}}-\frac{e^{y}}{m_{\phi}^{2}}+\frac{R_{\chi}e^{-y}}{m_{\phi}}+\frac{R_{\chi}e^{y}}{m_{\phi}}\right)^{2}.
\end{eqnarray}
\addcontentsline{toc}{section}{References}
\bibliographystyle{JHEP}
\bibliography{note}

\end{document}